\documentclass[12pt,preprint]{aastex}
\usepackage{natbib}

\slugcomment{ApJ Letters, accepted}
 
\shorttitle{[CII] line emission in BR1202-0725 at $z=4.7$}
\shortauthors{Wagg et al.}
 
\begin{document}

\title{[CII]  line emission in massive star-forming galaxies at $z=4.7$}

\author{J.~Wagg\altaffilmark{1}, T.~Wiklind\altaffilmark{2}, C.~L.~Carilli\altaffilmark{3,4}, D.~Espada\altaffilmark{5},  A.~B.~Peck\altaffilmark{2,6}, D.~Riechers\altaffilmark{7}, F.~Walter\altaffilmark{6,8}, A.~Wootten\altaffilmark{6}, M.~Aravena\altaffilmark{1}, D.~Barkats\altaffilmark{2}, J.~R.~Cortes\altaffilmark{2}, R.~Hills\altaffilmark{2}, J.~Hodge\altaffilmark{2,8}, C.~M~V.~Impellizzeri\altaffilmark{2,6}, D.~Iono\altaffilmark{5}, A.~Leroy\altaffilmark{6}, S.~Mart\'in\altaffilmark{1}, M.~G.~Rawlings\altaffilmark{2}, R.~Maiolino\altaffilmark{4}, 
R.~G.~McMahon\altaffilmark{9}, K.~S.~Scott\altaffilmark{6}, E.~Villard\altaffilmark{2}, C.~Vlahakis\altaffilmark{2}}

\email{jwagg@eso.org}
\altaffiltext{1}{European Southern Observatory, Casilla 19001, Santiago, Chile}
\altaffiltext{2}{Joint ALMA Observatory, Alonso de Cordova 3107, Vitacura, Santiago, Chile}
\altaffiltext{3}{National Radio Astronomy Observatory, Socorro, USA and Visiting Fellow, Cavendish Laboratory, Cambridge, UK}
\altaffiltext{4}{Cavendish Laboratory, University of Cambridge, Cambridge UK}
\altaffiltext{5}{National Astronomical Observatory of Japan, Tokyo, Japan}
\altaffiltext{6}{National Radio Astronomy Observatory, Charlottesville, USA}
\altaffiltext{7}{Dept. of Astronomy, California Institute of Technology, Pasadena, USA}
\altaffiltext{8}{Max-Planck Institute for Astronomy, Heidelberg, Germany}
\altaffiltext{9}{Institute of Astronomy, University of Cambridge, Cambridge, UK}

\begin{abstract}
\noindent 

We present Atacama Large Millimeter/submillimeter 
Array (ALMA) observations of the [CII] 157.7$\mu$m fine structure line and
thermal dust continuum emission from a pair of gas-rich 
galaxies at $z=4.7$, BR1202-0725.  This system consists of a luminous quasar host galaxy and a 
bright submm galaxy (SMG), while a fainter star-forming galaxy is also spatially coincident within a $4''$ (25\,kpc) region.  All three galaxies are detected in the submm
continuum, indicating FIR luminosities in excess of 10$^{13}$~L$_{\odot}$ for the two most luminous objects. The SMG and the quasar host galaxy are both detected in [CII] line emission with luminosities, $L_{\rm [CII]} = (10.0\pm 1.5)\times10^{9}$~L$_{\odot}$ and $L_{\rm [CII]} = (6.5\pm 1.0)\times10^{9}$~L$_{\odot}$, respectively. We estimate a luminosity ratio, $L_{\rm [CII]}/L_{\rm FIR} = (8.3\pm 1.2)\times 10^{-4}$ for the starburst SMG to the North, and $L_{\rm [CII]}/L_{\rm FIR} = (2.5\pm 0.4)\times 10^{-4}$ for the quasar host galaxy, in agreement with previous high-redshift studies that suggest lower [CII]-to-FIR luminosity ratios in quasars than in starburst galaxies. The third fainter object with a flux density, $S_{340GHz} = 1.9\pm0.3$~mJy, is coincident with a Ly-$\alpha$ emitter and is detected in  \textit{HST} ACS~F775W and F814W images but has no clear counterpart in the \textit{H}-band. Even if this third companion does not lie at a similar redshift to BR1202-0725, the quasar and the SMG represent an overdensity of massive, infrared luminous star-forming galaxies within 1.3~Gyr of the Big Bang.

\end{abstract}


\section{Introduction}

The formation of massive galaxies at high-redshift is often traced by intense emission at observed submm and mm-wavelengths, an indication of dust heated by star-formation or an active galactic nucleus (AGN). The observed populations of submm-selected starburst galaxies (SMGs; e.g. Blain et al.\ 2002) and FIR luminous quasar host galaxies (e.g. Omont et al.\ 1996; Wang et al.\ 2008) likely represent the formation of the most massive galaxies at high-redshift given their extreme FIR luminosities\footnote{We define FIR luminosity as L$_{\mathrm{FIR}}$ integrated over restframe 40--500$\mu$m.}, 
$L_{\rm FIR} \sim 10^{13}$ L$_\odot$. About one third of optically luminous quasar host galaxies at $z \ga 5$ display such high FIR luminosities (e.g. Wang et al.\ 2008), and the implied star-formation rates $> 10^3$~M$_\odot$~yr$^{-1}$ may be the main source of power. If these rates 
can be maintained for greater than 10$^7$~years, such objects could evolve into massive early-type galaxies with old stellar populations (e.g. Daddi et al.\ 2009).

BR1202-0725 was discovered at $z = 4.7$ with radio continuum emission resolved into two distinct components (Omont et al. 1996; Yun et al.\ 2000; Carilli et al.\ 2002), where the Southern 
emission is associated  with an optically luminous quasar, and the Northern a heavily obscured starburst submm galaxy (SMG). These are separated by 3.8$''$ (25\,kpc) in the North-West direction, and each is detected in interferometric imaging of the high-\textit{J} CO line emission  (Omont et al.\ 1996; Guilloteau et al.\ 1999). The  single-dish CO~\textit{J}=1-0 line luminosity indicates a combined mass of $\sim$8$\times$10$^{10}$~M$_{\odot}$ in cold molecular gas (Riechers et al.\ 2006). The difference in the high-\textit{J} CO line redshifts and line profiles demonstrate that BR1202-0725 is not gravitationally lensed, but rather the ongoing merger between two gas-rich galaxies. Submm and mm-wavelength imaging with the Submillimeter Array (SMA) and the Plateau de Bure Interferometer (PdBI) resolves the thermal dust continuum emission into the Northern and Southern components (Omont et al.\ 1996; Iono et al.\ 2006), revealing extreme FIR luminosities and an estimated star-formation rate $> 1000$~M$_{\odot}$~yr$^{-1}$ for each component. The likely interaction between two massive galaxies with such intense multi-wavelength emission makes this an ideal system for studying the formation of galaxies in the young Universe.

The brightest emission line in the interstellar medium (ISM) of galaxies is typically the 157.74~$\mu$m [CII] line,  arising primarily in photodissociation regions (PDRs) and the cold neutral medium. 
Its intensity and ubiquitous nature make this line an ideal means to study the formation and kinematics of high-redshift galaxies.  The [CII] line luminosity typically contains $\sim$0.1--1\% of the total far-infrared (FIR) luminosity in quiescent galaxies, and less than 0.1\% in ultra-luminous infrared galaxies (ULIRGs; Crawford et al.\ 1985; Stacey et al.\ 1991; 
 Wright et al.\ 1991; Malhotra et al.\ 1997). For galaxies at redshifts, $z \ga 1$, this line can be redshifted into the submm-wavelength atmospheric windows and has already been detected in more than ten high-redshift objects (e.g. Maiolino et al.\ 2005, 2009; Walter et al.\ 2009;  Ivison et al.\ 2010; Stacey et al.\ 2010; Wagg et al.\ 2010; De~Breuck et al.\ 2011), including an SMA detection in the SMG BR1202-0725 North (Iono et al.\ 2006).

In this work, we present observations of the thermal dust continuum and redshifted [CII] line emission in the quasar host galaxy, BR1202-0725 at $z = 4.7$, obtained using a limited number of ALMA antennas during commissioning and science verification. Throughout this {\it Letter}, we adopt a cosmological  model with $(\Omega_\Lambda, \Omega_m, h) = (0.73, 0.27, 0.71)$ (Spergel et al.\ 2007).

\section{Observations}

The BR1202-0725 system was observed at 335~GHz (Band 7) with 17 12-m diameter antennas of the Atacama Large Millimeter/submillimeter Array as part of science verification on January 16, 2012.  Most of the antennas were in a compact configuration with two antennas on longer baselines, providing a maximum baseline length of $\sim$350-m. The total on-source observing time was 25 minutes during which the target was above 60$^\circ$ elevation. Even with such a short integration time, ALMA is already an order of magnitude more sensitive than any previous submm observations of the system. The total bandwidth of these observations was 7.5~GHz, divided into four spectral windows of width 1.875~GHz. The correlator channel spacing was 15.6~MHz, corresponding to $\sim$14~km~s$^{-1}$ at the observed frequency of 334.96~GHz. The rest frequency of the [CII] line is 1900.539~GHz. The data presented here lack measurements of system temperature and water vapor radiometer correction which are part of the standard ALMA observing strategy, however earlier observations made during the same program (on Jan 14, 2012) contain full calibration information.  As the earlier observations were made while the target was at lower elevation, the rms noise level was higher and so the data are not included in this paper. However, these observations were used to verify the flux scale of the data presented here.

 Data analyis was performed using the \textit{CASA} software package\footnote{http://casa.nrao.edu}. 
The flux density scale was derived using observations of Titan. We derive a flux density of 17~Jy for the complex gain calibrator, 3C279, located 13$^\circ$ from BR1202-0725. Some uncertainty in the flux density of 3C279  is present due to the lack of system temperature measurements with these data. We estimate this uncertainty to be 15\% and include this in our measured flux densities and integrated intensities. A 7~min calibration cycle  was used and the weather conditions were dry (precipitable water vapour $\sim 0.7$~mm). The gains measured for 3C279 were stable to within a few degrees and a  few percent during the observation. The spectral bandpass shape was also measured using 3C279. 

The sensitivity is high enough to allow self-calibration using the  line-free emission from BR1202-0275. Amplitude and phase self-calibration were done using a 3~min averaging time for complex gain solutions, resulting in a higher dynamic range in both the line and continuum images. A continuum map of BR1202-0725 was made by imaging the last three spectral windows, which results in a 340~GHz continuum sensitivity of 0.2~mJy per beam. A UV plane model for the continuum emission is derived from these spectral windows and subtracted from the first window. After continuum subtraction and resampling to $\sim$28~km~s$^{-1}$ resolution, the channel rms is 2.0~mJy per beam.

\section{Results and analysis}

\begin{table*}[ht]
\centering
\caption{Observed continuum and [CII] line parameters for BR1202-0725.}
\begin{center}
\begin{tabular}{l c c c c c c}
\hline \hline
Source  & $S_{340GHz}$ & $z _{\rm [CII]}$$^a$ &  $\Delta V_{FWHM}$$^a$     & $SdV$ & $L_{\rm [CII]}$  &   $L_{\rm [CII]} / L_{FIR}$  \\
             &    [mJy]         &                &   [km~s$^{-1}$]  &   [Jy~km~s$^{-1}$] &  $\times10^9$ [L$_{\odot}$] & $\times 10^{-4}$ \\ 
\hline 
BR1202-0725 North  &  18.8$\pm$2.8  &   4.6906    &   722$\pm$12  &   14.7$\pm$2.2   & 10.0$\pm$1.5 & $8.3\pm1.2$ \\
BR1202-0725 South  &  18.0$\pm$2.7  &   4.6943    &    328$\pm$6   &   9.6$\pm$1.5  &   6.5$\pm$1.0  & $2.5\pm0.4$ \\
BR1202-0725 SW   &   1.9$\pm$0.3  &       -     &      -        &     -      &   -    & -  \\
\hline
\end{tabular}
\vskip 0.1in
\noindent $^a${} The [CII] redshift and line width ($\Delta V_{FWHM}$) are determined from the best fit parameters of a Gaussian fit to the spectra. \\
\label{ref:table1}
\end{center}
\end{table*}

\vskip 0.1in
The synthesized beamsize of the 340~GHz continuum observations is $1.30'' \times 0.86''$, corresponding to physical scales of $8.6 \times 5.7$~kpc at $z = 4.7$. The total 340~GHz flux density of the system is $43\pm 6$~mJy (including calibration uncertainties), in excellent agreement with the previous $\sim$14$''$ resolution 850~$\mu$m SCUBA observations ($S_{850\mu m} = 42\pm 2$~mJy; McMahon et al.\ 1999). Figure~1 shows a map of the 340~GHz continuum emission in BR1202-0725, while Figure~2 shows this thermal dust continuum emission overlaid on the \textit{HST} ACS F775W, WFPC2 F814W and NIC2 F160W images. Both the quasar host galaxy to the South, and the optically obscured starburst SMG to the North are clearly detected at 340~GHz (Table~1). Gaussian fitting of the continuum emission in the image plane suggests that neither of these sources is resolved. A third faint source is detected in 340~GHz continuum emission to the South-West of the quasar with a flux density $S_{340GHz} = 1.9\pm0.3$~mJy. We hereafter refer to this source as BR1202-0725~SW, and its faintness places it below the detection threshold of previous interferometric observations at these frequencies (Iono et al.\ 2006).

For the quasar and the SMG we adopt the FIR luminosity estimates of Iono et al.\ (2006), who fit the FIR-to-radio observations using the templates of Yun \& Carilli (2002) and also measure similar continuum flux densities for the two components at these frequencies. For BR1202-0725 North, they estimate $L_{\rm FIR} = 1.2\times 10^{13}$~L$_{\odot}$, and $L_{\rm FIR} = 2.6\times 10^{13}$~L$_{\odot}$ for BR1202-0725 South. Using our measurements of the 340~GHz continuum flux densities and assuming $\beta = 1.5$ with dust temperatures of 35~K for the SMG (e.g. Kovacs et al.\ 2006) and 60~K for the quasar (e.g. Beelen et al.\ 2006; Weiss et al.\ 2007), we estimate similar FIR luminosities to those estimated by Iono et al.\ (2006) and simultaneously reproduce the combined 350~$\mu$m flux density of the pair (Benford et al.\ 1999).
BR1202-0725 SW is not detected in the 1.4~GHz observations of Carilli et al.\ (2002), implying a 3-$\sigma$ upper-limit of 48~$\mu$Jy at 1.4~GHz. The dust temperature is uncertain, and so to estimate the FIR lumosity of BR1202-0725 SW we scale down the luminosities of the two bright sources by the ratio of the flux densities, implying $L_{\rm FIR} \sim 10^{12}$~L$_{\odot}$. Similar  luminosities are estimated for BzK-selected star-forming galaxies and dust obscured galaxies (DOGs) at $z \sim 2$ (e.g. Pope et al.\ 2008; Daddi et al.\ 2008) and we might expect comparable 350~GHz flux densities to arise from a range of different high-redshift galaxy populations.

From the spectral line data cubes we extract spectra at the positions of peak intensity in the submm image. Figure~3 shows the spectra of the three continuum sources where redshifted [CII] line emission is clearly detected in both the quasar host galaxy and the SMG. The peak in the [CII] line emission from both objects is spatially coincident with the FIR continuum and previously observed CO~\textit{J=}2-1 line emission (Carilli et al.\ 2002). There is also a hint of line emission at the red end of the spectrum observed toward BR1202-0725 SW, although we cannot exclude the possibility that this emission is spurious.
We fit Gaussian line profiles to the two detected [CII] lines and the results are given in Table~1. The line width measured in the quasar host galaxy ($\Delta V_{FWHM} \sim 330$~km~s$^{-1}$) is a factor of $\sim$2 smaller than that observed in the SMG ($\Delta V_{FWHM} \sim 720$~km~s$^{-1}$), which is in broad agreement with the difference observed in CO emission lines detected in these two populations at high-redshift (e.g. Solomon \& Vanden~Bout 2005; Carilli \& Wang 2006). The line widths are also in excellent agreement with previous observations of CO line emission (Omont et al.\ 1996; Carilli et al.\ 2002). The [CII] linewidth and redshift of BR1202-0725~North are in agreement with the previous SMA observations of Iono et al.\ (2006).

Following the definition of line luminosity  (see Solomon \& Vanden~Bout 2005), we calculate $L_{\rm [CII]} = (10.0\pm1.5)\times10^{9}$~L$_{\odot}$ and $L_{\rm [CII]} = (6.5\pm1.0)\times10^{9}$~L$_{\odot}$ for BR1202-0725 North (SMG) and South, respectively. Combined with our estimates of the FIR luminosity, we calculate [CII]-to-FIR luminosity ratios $L_{\rm [CII]}/L_{\rm FIR} = (8.3\pm 1.2)\times 10^{-4}$ for the SMG, and $L_{\rm [CII]}/L_{\rm FIR} = (2.5\pm 0.4)\times 10^{-4}$ for the quasar host galaxy. These ratios are in agreement with that found in other high-redshift galaxies observed in [CII] line emission (e.g. Stacey et al.\ 2010; Wagg et al.\ 2010; Cox et al.\ 2011).

\section{Discussion}

The extreme FIR luminosities observed in BR1202-0725 North and South would imply obscured star-formation rates in excess of 1000~M$_{\odot}$~yr$^{-1}$ (e.g. Kennicutt 1998). The SFR for the fainter SW component is $\sim$10 times lower due its lower FIR luminosity. As such, this system likely represents the build up of an extremely massive galaxy, possibly about to undergo a major merger. \textit{Chandra} detects X-ray emission in BR1202-0725 South and tentatively in BR1202-0725 North (Iono et al.\ 2006), suggesting that the rapid build-up of the stellar mass in these objects is coeval with the growth of a supermassive black hole (e.g. Hopkins et al.\ 2006).

Mounting evidence for the presence of large-scale outflows arising from AGN has emerged from the detections of large-scale emission from ionized gas in high-redshift quasars (e.g. Cano-D{\'{\i}}az et al.\ 2012) and the presence of P-Cygni profiles and extended wings in the molecular  lines observed in nearby ultraluminous infrared galaxies  (ULIRGs; e.g. Sakamoto et al.\ 2009; Feruglio et al.\ 2010; Aalto et al.\ 2012; Sturm et al.\ 2012). The [CII] line profile of the quasar BR1202-0725 South, is described very well by a Gaussian profile, however a low intensity, broad line width extension is seen toward positive velocities. If all of the ionized gas in this galaxy were to arise from a region of gas rotating in virial equilibrium around the AGN, then the excess emission seen in the [CII] line may be due to a single-sided outflow of ionized gas from the quasar. Such broad [CII] line wings have been detected by \textit{Herschel} in Mrk231 (Fischer et al.\ 2010) and by the PdBI in the z=6.4 quasar host galaxy J1148+5251 (Maiolino et al.\ 2012).
Higher angular resolution imaging is required to confirm the presence of an outflow in BR1202-0725 South.

The luminosity ratio between emission from the [CII] line and the FIR continuum provides constraints on the physical conditions of the interstellar gas in galaxies.  The [CII] line typically exhibits a luminosity $\sim$0.1--1\% that of the total FIR luminosity in quiescent galaxies, and less than 0.1\% in ultra-luminous infrared galaxies (ULIRGs; Crawford et al.\ 1985; Stacey et al.\ 1991; 
 Wright et al.\ 1991; Malhotra et al.\ 1997). This ratio may also be lower in luminous quasars than in starburst galaxies (e.g. Stacey et al.\ 2010), which is supported by the observations presented here.  One explanation for the decrease in the [CII]-to-FIR ratio in ultraluminous infrared galaxies is an increase in the rate of UV photons absorbed by dust grains, which decreases the number of photons available to ionize the gas (e.g. Abel et al.\ 2009; Papadopoulos et al.\ 2010). Graci\'a-Carpio et al.\ (2011) find that the scatter in the [CII]-to-FIR ratio can be decreased if one plots this ratio as a function of the FIR luminosity weighted by the molecular gas mass. In the case of BR1202-0725 the CO~\textit{J}=2-1 line luminosity (or the cold molecular gas mass) is the same in both the Northern and Southern sources (Carilli et al.\ 2002), and so weighting the FIR luminosity by the molecular gas mass would have an equivalent impact on each. Similarly high signal-to-noise [CII] line spectra and FIR continuum observations of significant samples of high-redshift quasars and starburst galaxies will reveal the typical structure and kinematics of the ionized gas in these objects, and finally resolve the origin of the observed line to continuum ratios.

The presence of two SMGs in the BR1202-0725 system, potentially situated within 25\,kpc of the QSO would likely represent an extreme merger event at high-redshift. The two brightest submm components, one of which is the QSO host galaxy, contain large amounts of molecular gas ($>10^{10}$\,M$_{\odot}$) and each component is likely to have a total mass (gas $+$ stars) in excess of $10^{11}$\,M$_{\odot}$, similar to comparably luminous SMGs at these redshifts (e.g. Riechers et al.\ 2010). Giallongo et al.\ (1998) searched for the presence of an overdensity of galaxies in the vicinity of BR1202-0725, finding no evidence for an increase in the number density of optically-selected galaxies at these redshifts. However, our deep continuum observations with ALMA suggest that the star-formation in galaxies associated with this field is heavily obscured. There are two Ly-$\alpha$ detected galaxies in the vicinity of BR1202-0725 (Hu et al. 1996, 1997). The brightest Ly-$\alpha$ emitter is associated with the NW SMG, while a fainter Ly-$\alpha$ emitter is associated with the SW component. Both of these galaxies are detected in the HST F775W and F814W images, but only the brighter NW galaxy is seen in the NICMOS F160W image (Figure~2).

The FIR luminosities of the three submm components implies a combined SFR of several $10^3$\,M$_{\odot}$\,yr$^{-1}$. Several observational facts suggest that star formation is indeed powering the high luminosities. The rest-frame UV light associated with the NW component has been shown to be the result of star formation rather then reflected QSO light (Fontana et al. 1998). The combination of high inferred star formation rates and significant [CII] line emission is consistent with a scenario whereby this line emission originates in PDR regions associated with star formation. Taken together, it implies that star formation, possibly triggered by gravitational interaction, rather than processes associated with the QSO are powering the UV as well as the FIR emission. 

 Daddi et al.\ (2009) estimate that the number density of submm luminous starburst galaxies in the redshift range, $3.5 < z < 6.0$, is 10$^{-5}$~Mpc$^{-3}$. Two such galaxies with likely masses $\ga 10^{11}$\,M$_{\odot}$ found within a projected radius of 25\,kpc therefore represents a significant over-density at $z=4.7$, possibly associated with a `proto-cluster' region when the Universe was only 1.3\,Gyr old. The most probable descendant of this interaction will be a  massive galaxy (several 10$^{11}$\,M$_{\odot}$), with very little gas and dust remaining. Massive and evolved galaxies have been found at high redshifts in several deep surveys (e.g. Cimatti et al.\ 2004) and could represent the result of similarly violent merging events as we are currently witnessing in the BR1202-0725 system.

\section{Summary}
\label{sec:sum}

We present the first ALMA observations of redshifted [CII] line and thermal dust continuum emission in high-redshift quasar host galaxies, observing BR1202-0725 at $z = 4.7$ as part of commissioning and science verification. Our continuum image reveals a faint submm companion to the previously known quasar host galaxy and FIR luminous SMG, however the redshift of this object is unknown. We also measure high signal-to-noise detections of the ionized carbon line in both the Northern starburst galaxy and the quasar host galaxy to the South. The line widths agree with previous observations of high-\textit{J} CO line emission, with a narrower line profile observed in the quasar. The decrease in the [CII]-to-FIR ratio measured in nearby and high-redshift ULIRGs is also observed in BR1202-0725, with a higher ratio in the starburst galaxy, in agreement with surveys of quasars and starburst galaxies at $z \sim 1-2$ (Hailey-Dunsheath et al.\ 2010; Stacey et al.\ 2010). Tentative evidence for an outflow of ionized gas from the quasar is observed in the [CII] line spectrum, and higher angular resolution observations of this object with ALMA will allow us to constrain the spatial extent of this gas and measure the total energy and kinematics.

\acknowledgements We thank the referee for helpful suggestions based on the original manuscript. This work was co-funded under the Marie Curie Actions of the European Commission (FP7-COFUND). We thank all those involved in the ALMA project for making these observations possible. This paper makes use of the following ALMA data: ADS/JAO.ALMA\#2011.0.00006.SV. ALMA is a partnership of ESO (representing its member states), NSF (USA) and NINS (Japan), together with NRC (Canada) and NSC and ASIAA 
(Taiwan), in cooperation with the Republic of Chile. The Joint ALMA 
Observatory is operated by ESO, AUI/NRAO and NAOJ. DR acknowledges support from NASA through a Spitzer Space Telescope grant.

\begin{figure}
\epsscale{0.9}
\plotone{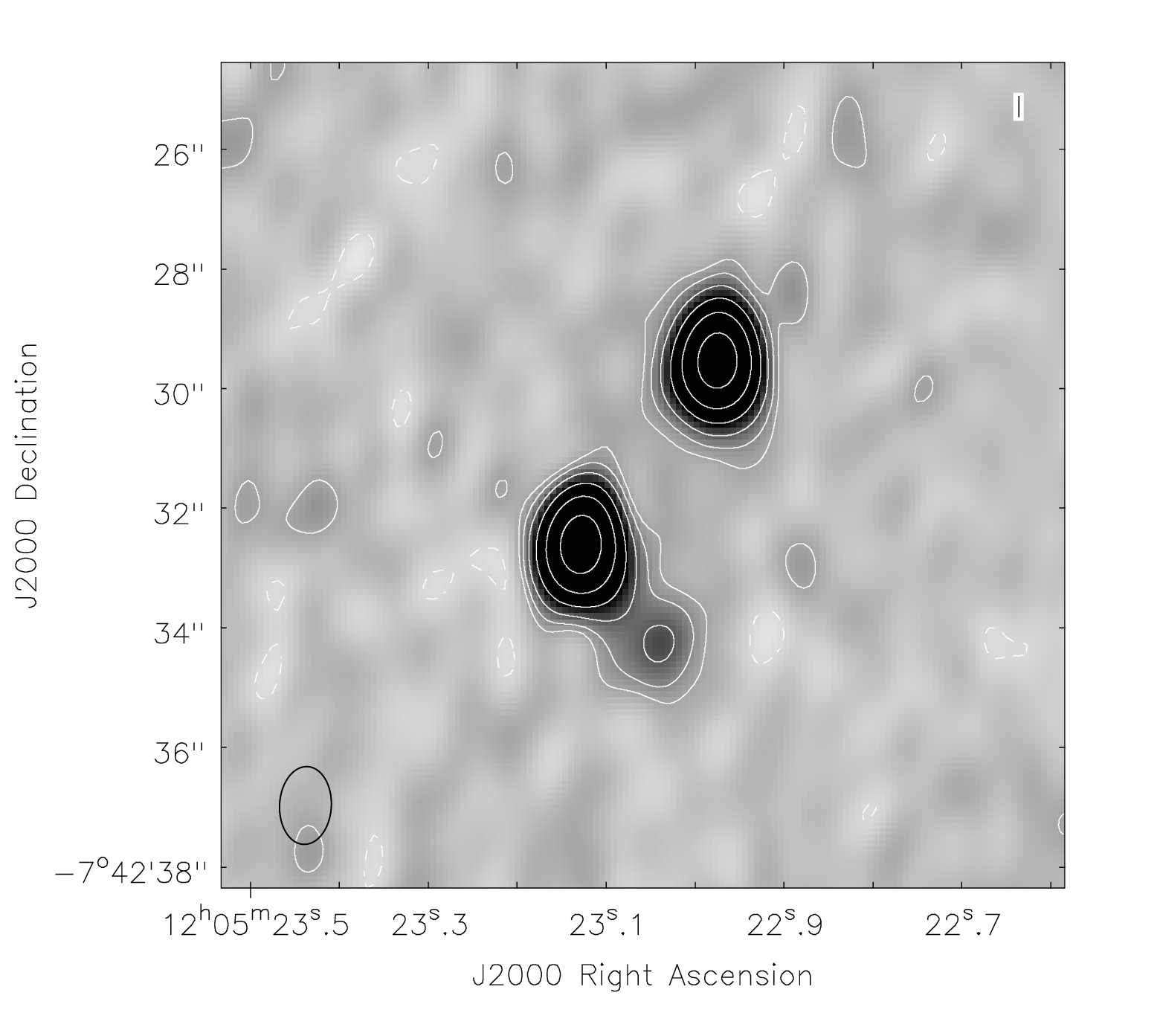}
\caption{
340~GHz continuum image of the BR1202-0725 system at $z = 4.7$ generated from the line-free spectral windows. The synthesized beamsize is $1.30'' \times 0.86''$, and the continuum rms is 0.2~mJy per beam (not including calibration uncertainties). Contour intervals are (-2, 2, 4, 8, 16, 32 and 64)$\times \sigma$. The quasar host galaxy to the South and the SMG to the North are detected at high significance, while the faint continuum source we refer to as BR1202-0725~SW is detected with a significance of 9.5-$\sigma$. 
}
\end{figure}

\begin{figure}
\epsscale{1.}
\plotone{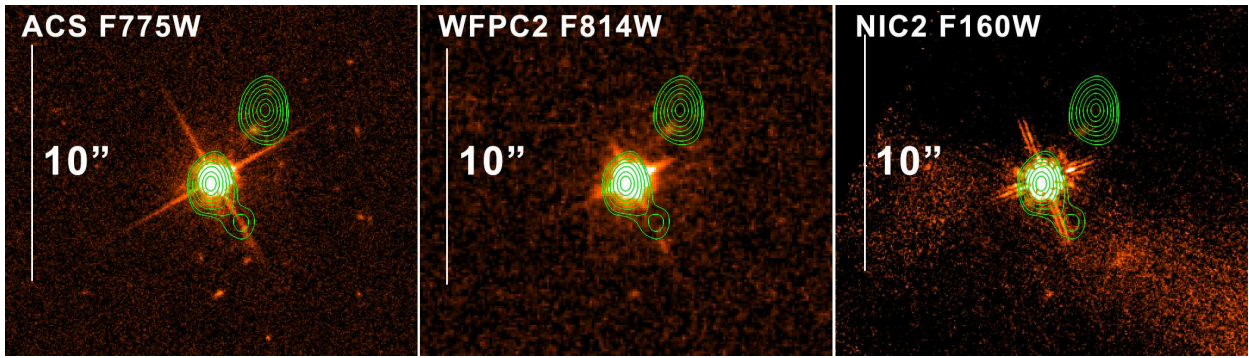}
\caption{Archival \textit{HST} ACS~F775W (prop ID 10417), WFPC2~F814W (prop ID 5975) and NICMOS F160W (prop ID 7266) images of the BR1202-0725 system at $z = 4.69$. Contour levels show the intensity of the 340~GHz continuum emission measured by ALMA, where the lowest contour level is 0.9~mJy~per~beam. The synthesized beamsize is $1.30'' \times 0.86''$. 
The QSO BR1202-0725 is clearly visible in all three \textit{HST} images. The two Ly-$\alpha$ galaxies detected by Hu et al. (1996) can be seen as faint emission in the F775W and F814W images. Only the NW Ly-$\alpha$ galaxy is seen in the F160W image. Both the NW and SW Ly-$\alpha$ galaxies are offset from the peaks in the submm continuum emission,   and are apparently extended toward the QSO.
}
\end{figure}

\begin{figure}
\epsscale{0.4}

\plotone{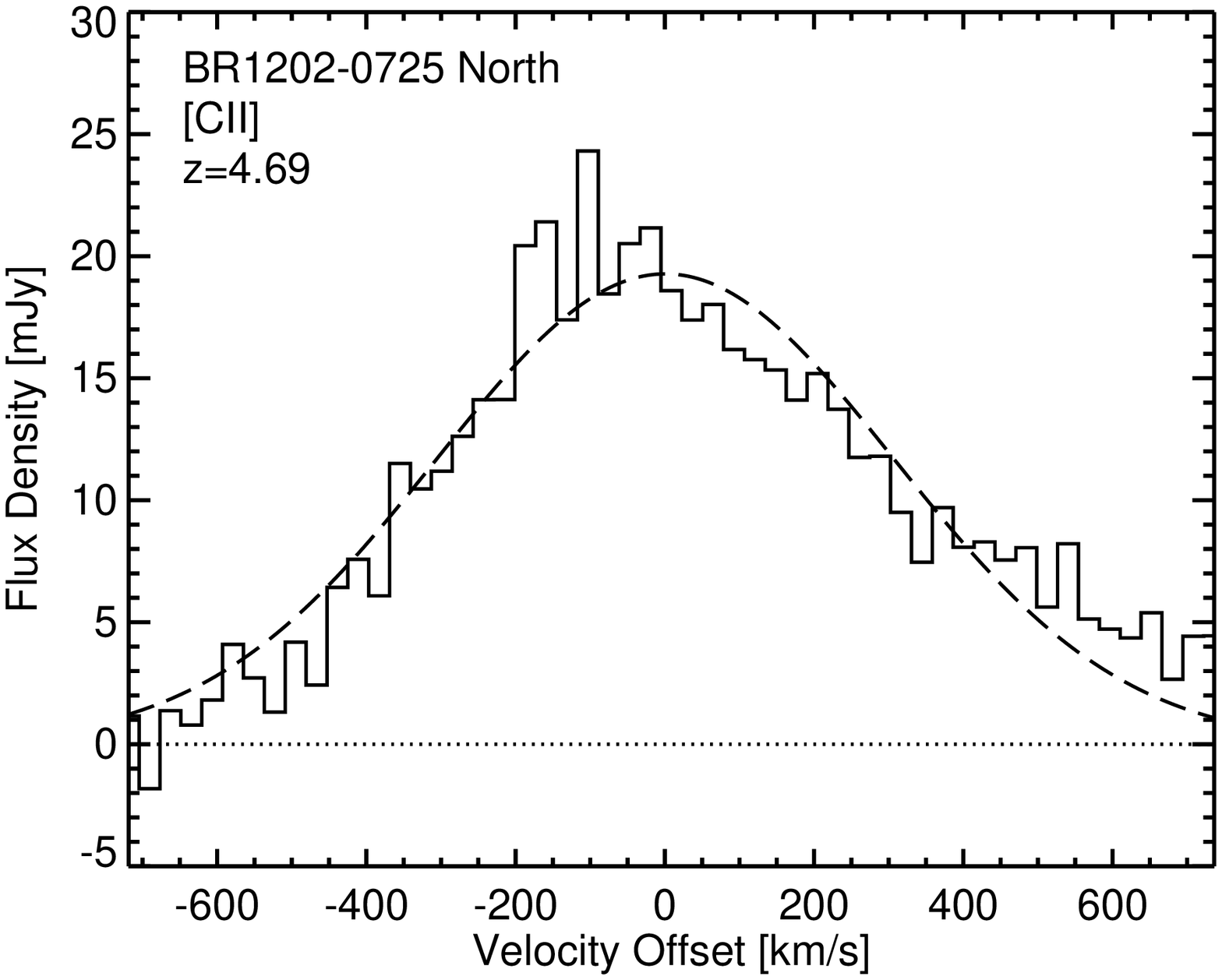}

\plotone{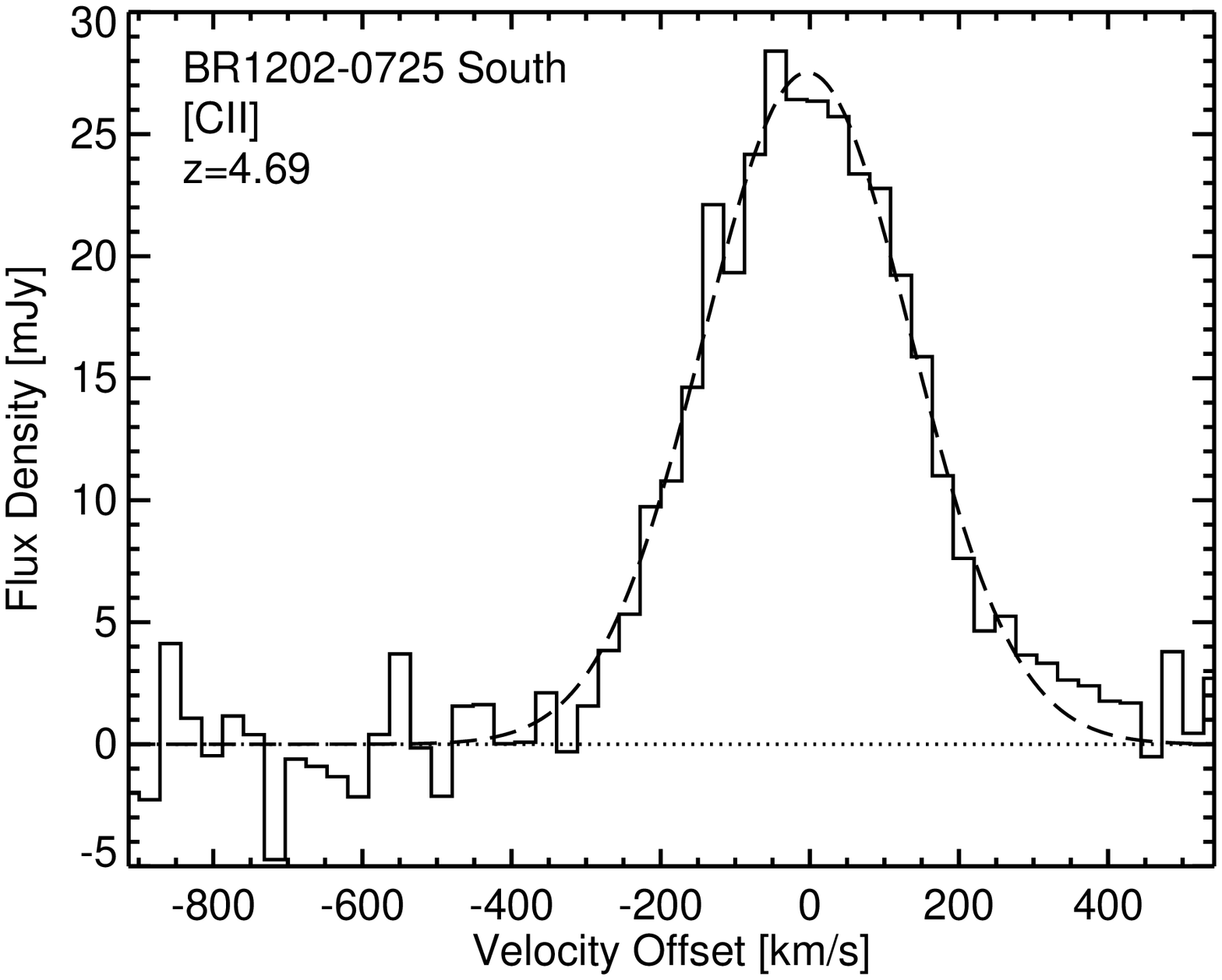}

\plotone{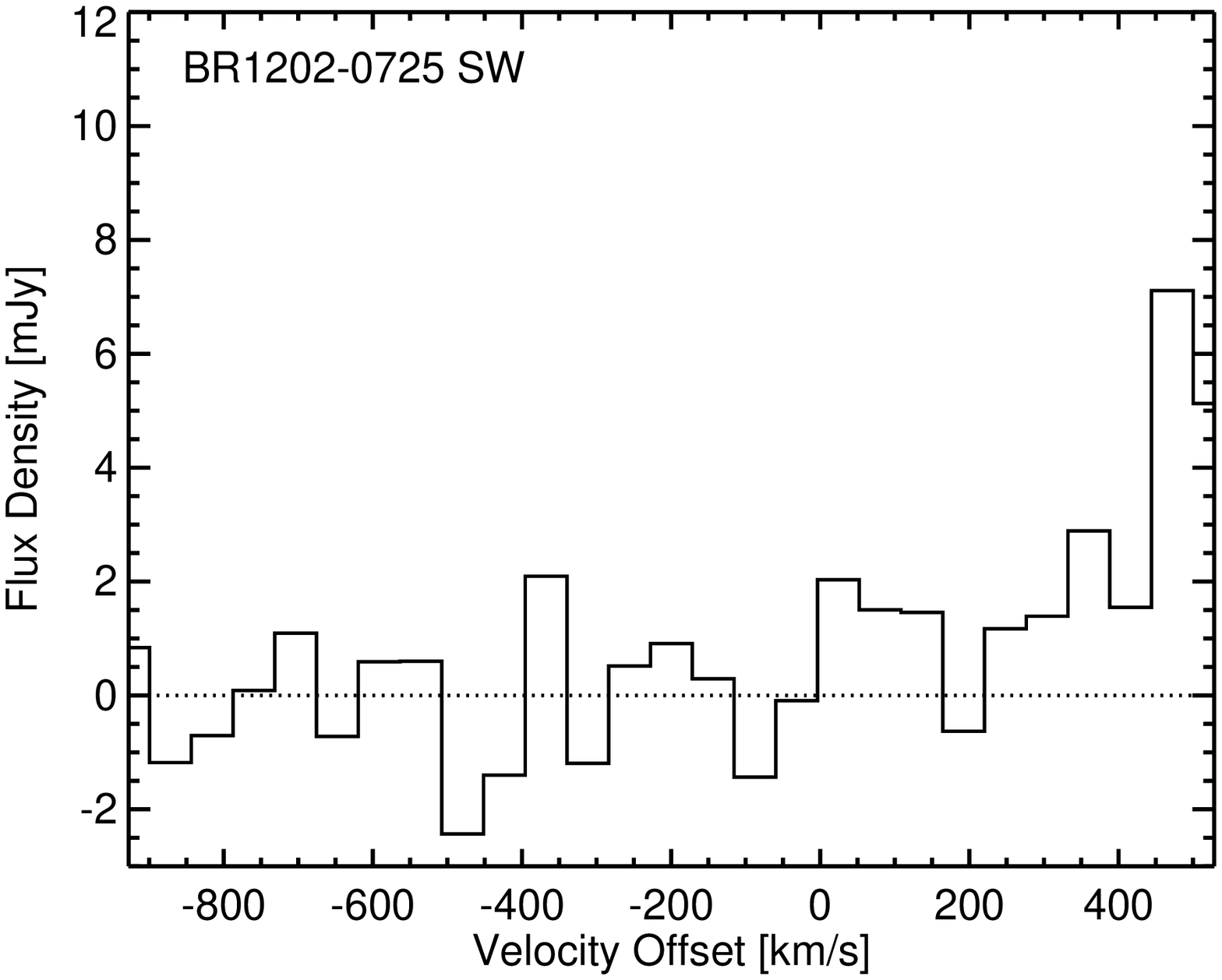}
\caption{Continuum subtracted spectra taken at the peak intensity of the three submm sources. For BR1202-0725 North and South, the spectra are sampled at $\sim$28~km~s$^{-1}$ resolution, and we have fit the [CII] line emission using Gaussian profiles (\textit{dashed} lines) and the best-fit parameters are given in Table~1. 
In the BR1202-0725 North spectrum (\textit{top}), the velocity scale has been calculated with respect to $z = 4.6906$ for the [CII] line, while the spectra of BR1202-0725 South (\textit{middle}) is plotted with respect to $z = 4.6943$. 
 The spectrum of BR1202-0725 SW is plotted in the \textit{bottom} panel and the zero velocity scale has been set to the same as BR1202-0725 South ($z = 4.6943$ for the [CII] line), although the redshift of this object is unknown. The rms is 1.3~mJy per $\sim$56~km~s$^{-1}$ channel in the spectrum of BR1202-0725 SW. 
}
\end{figure}

\end{document}